\documentclass[preprint,12pt]{elsarticle}
\usepackage{amssymb}

\usepackage{color, fullpage}
\renewcommand\baselinestretch{1.12}

\journal{Elsevier}

\newcommand{\bA}{\mathbf{A}}
\newcommand{\bF}{\mathbf{F}}
\newcommand{\bU}{\mathbf{U}}

\newdefinition{rmk}{Remark}

\begin{document}

\begin{frontmatter}

\title{MIXING LAYER AND TURBULENT JET FLOW \\ IN A HELE--SHAW CELL}

\author[label1,label2]{Alexander Chesnokov}
\ead{Corresponding author: chesnokov@hydro.nsc.ru}
\author[label1,label2]{Valery Liapidevskii} 
\ead{liapid@hydro.nsc.ru}

\address[label1]{Lavrentyev Institute of Hydrodynamics SB RAS, 15 Lavrentyev Ave., Novosibirsk 630090, Russia}
\address[label2]{Novosibirsk State University, 1 Pirogova Str., Novosibirsk 630090, Russia}

\begin{abstract}
A plane turbulent mixing in a shear flow of an ideal homogeneous fluid confined between two relatively close rigid walls is considered. The character of the flow is determined by interaction of vortices arising at the nonlinear stage of the Kelvin--Helmholtz instability development and by turbulent friction. In the framework of the shallow water theory and a three-layer representation of the flow, one-dimensional models of a mixing layer are proposed. The obtained equations allow one to determine averaged boundaries of the region of intense fluid mixing. Stationary solutions of the governing equations are constructed and analysed. Using the averaged flow characteristics obtained by one-dimensional equations, a hyperbolic system for determining the velocity profile and Reynolds shear stress across the mixing layer is derived. Comparison with the experimental results of the evolution of turbulent jet flows in a Hele--Shaw cell shows that the proposed models provide a fairly accurate description of the average boundaries of the region of intense mixing, as well as the velocity profile and Reynolds shear stress across the mixing layer.
\end{abstract}

\begin{keyword}   
shallow water equations; mixing layer; jet flow; Hele--Shaw cell
\end{keyword}

\end{frontmatter}

\section{Introduction} 

The interaction of fast and slow flows when they merge leads to the appearance of vortex structures and the formation of a mixing layer~\cite{Brown_1974, Streeter_1998, Ho_1984}. Interest in modelling of such flows is associated with geophysical and technical applications, in which there is a need to describe the mixing processes during the confluence of rivers or streams in open and closed channels. For a wide class of flows, the study of shear instability can be carried out in the framework of the two-dimensional shallow water theory, taking into account the effect of bottom friction. In this case, the character of the quasi-two-dimensional flow and the mixing intensity are determined by interaction of vortices arising at the nonlinear stage of the Kelvin--Helmholtz instability and by friction caused by a small thickness of the fluid layer. Note that the fluid motion between two relatively close rigid walls can be interpreted as a flow in a Hele--Shaw cell with the law of friction on the walls corresponding to the turbulent flow regime. The study of such shear flows becomes relevant in the problems of underground hydrodynamics in modelling hydraulic fracturing.

Laboratory experiments on the formation and evolution of plane turbulent mixing layer for free surface flows were performed in \cite{Chu_1988, Uijttewaal_2000}. These results allowed one to obtain empirical formulas for estimating the size of the mixing area as well as determine the parameters at which the flow is stabilized and the growth of the intermediate mixing layer stops. In \cite{Sukhodolov_2010, Uijttewaal_2014} the field observations of confluent streams were discussed and compared with some theoretical results. Mathematical models of the plane mixing layer development based on the averaged equations of the shallow water theory were presented and verified in \cite{Booij_2001, Prooijen_2002}. The characteristic features of the development of mixing layers in shallow water at various flow parameters were numerically studied in \cite{Ghidaoui_2010, Kirkil_2015}. Theoretical and experimental study of quasi-two-dimensional processes of turbulent mixing in flows between rigid walls also attracts the attention of researchers. An experimental study and numerical simulation of the evolution of submerged jets in a Hele--Shaw cell were carried out in \cite{Shestakov_2015, Shestakov_2016}, where the formation of large vortex structures and meanders at Reynolds numbers of the order of ${\rm Re}=10^4$ were discussed. Here it was assumed that the width of the jet in the inlet section significantly exceeded the thickness of the channel. In works \cite{Landel_2012a, Landel_2012b}, on the contrary, flows are considered in which the width of the jet in the inlet section is smaller than the thickness of a Hele--Shaw cell. Experimental results on the meandering of turbulent jets were presented and a theoretical estimate for determining the averaged jet boundaries was given. The asymptotic theory of secondary steady flow in turbulent mixing layers was developed in \cite{Zametaev_2019}.

Improving the models of shear fluid flow is caused by the necessity of describing the nonlinear stage of the Kelvin--Helmholtz instability development and adequate modelling of a turbulent mixing in the framework of averaged equations of motion. In works~\cite{LT00, LCh_2014} an original method was proposed for constructing one-dimensional models for the propagation of nonlinear perturbations in the shear flow of a thin fluid layer. This method is based on the application of the theory of three-layer shallow water taking into account turbulent mixing in the intermediate layer. Implementation of the method allows for two different approaches. The first of them is based on averaging the Reynolds equation of the turbulent flow over the channel width with the additional assumption that the Reynolds shear stress is proportional to the specific kinetic energy of the fluctuating motion~\cite{Liap_2000, Liap_2019}. The second one uses the procedure proposed in \cite{Tesh_2007} for averaging shallow water equations for shear flows~\cite{Benney, ChL_2011}. It was applied to construct a three-layer model of horizontal mixing in fairly deep fluid flows in open channels~\cite{LCh_2016}. In \cite{ChL_2019}, this model was improved by taking into account bottom friction and verified by comparison with experimental data~\cite{Booij_2001} and the results of numerical simulations based on two-dimensional equations of shallow water theory. The same approach was successfully applied in \cite{GLCh16} to simulate the breaking of surface long waves taking into account the effects of vorticity and dispersion. Note that despite the differences in the derivation of 1D models of the mixing layer, the obtained equations of motion have a similar structure and, as a rule, give close quantitative results~\cite{LCh_2014}.

The purpose of this work is to derive, analyse and verify one-dimensional models describing the formation and evolution of the mixing layer in a shear flow of a homogeneous fluid confined between two relatively close parallel walls. Construction of the models is based on a three-layer representation of the flow taking into account turbulent mixing. In the next section, we derive 1D models for describing the mixing layer applying the Reynolds equations for turbulent flows (Model~I) and the shallow-water equations for shear flows (Model~II). Both models have a similar structure and allow a uniform presentation. The equations of motion include two empirical parameters responsible for mixing and dissipation of energy. In section~3, we construct and study stationary solutions. We show that the boundaries of the mixing layer obtained by Models I and II almost coincide. Then we show that the boundaries of the region of intense mixing for Hele--Shaw jet flows obtained from the proposed models are in good agreement with experimental data. Using the solution of the averaged 1D equations of motion, we derive a hyperbolic system for determining the velocity profile and Reynolds shear stress across the mixing layer. The results of calculating the velocity profile and shear stress are verified by comparison with the experimental data. In section~4, we modify these 1D models for the case of shear flows without a pressure gradient. Such models, in particular, are used to describe a submerged jet in an infinitely wide channel. A comparison of the calculation results utilizing stationary equations of motion with the available experimental data indicates the possibility of using these models to determine the averaged boundaries of a turbulent jet. Finally, we draw some conclusions. 

\section{Governing equations} 

In the shallow water approximation, a spatial flow of a homogeneous ideal fluid under a rigid-lid (in a Hele--Shaw cell) is described by the equations
\begin{equation} \label{eq:SW_2D}
  \begin{array}{l} \displaystyle
    U_t+UU_x+VU_y+p_x=-cU\sqrt{U^2+V^2}, \\[2mm]\displaystyle
    V_t+UV_x+VV_y+p_y=-cV\sqrt{U^2+V^2}, \\[2mm]\displaystyle
    U_x+V_y=0. 
  \end{array}
\end{equation}
Here $x$ and $y$ are the spatial variables; $t$ is the time; $U$ and $V$ are the velocity vector components; $p=\hat{p}/\rho$ is the ratio of the pressure $\hat{p}$ at the rigid-lid $z=h$ to the fluid density $\rho={\rm const}$; factor $c=c_f/h$ is the ratio of the friction coefficient $c_f={\rm const}$ to the channel thickness $h={\rm const}$. It supposes that in the $Oxy$ plane the flow region is bounded by the rigid walls $y=0$ and $y=Y$. Therefore, we supplement Eqs.~(\ref{eq:SW_2D}) with the boundary conditions
\begin{equation} \label{eq:BC}
   V|_{y=0}=0, \quad V|_{y=Y}=0. 
\end{equation}
The obvious consequence of system~(\ref{eq:SW_2D}) is the energy equation
\begin{equation} \label{eq:energy-gen}
  E_t+\big((E+p)U\big)_x+\big((E+p)V\big)_y=-c\big(U^2+V^2\big)^{3/2}, 
  \quad E=\frac{U^2+V^2}{2}\,.
\end{equation}  

\begin{figure}[t]
\begin{center}
\resizebox{.55\textwidth}{!}{\includegraphics{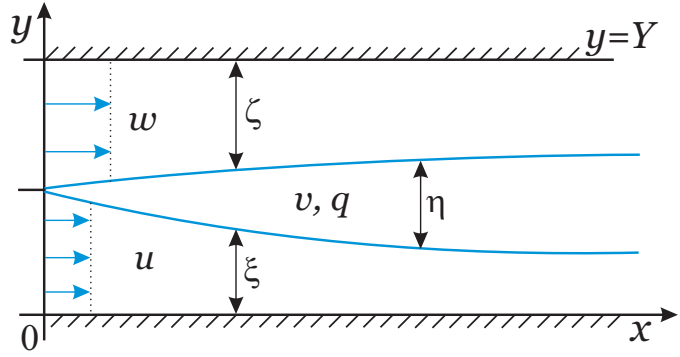}}\\[0pt]
{\caption{Three-layer representation of the flow (section by plane $z={\rm const}$).} \label{fig:fig_1}} 
\end{center}
\end{figure}

We consider the problem of the formation of a mixing layer in the confluence of two weakly shear flows. In the framework of the three-layer flow scheme (see Fig.~\ref{fig:fig_1}) we derive one-dimensional system of equations describing the evolution of the mixing layer. Let $\xi(t,x)$, $\zeta(t,x)$ and $\eta(t,x)$ denote the width of the outers and intermediate layers. The average flow velocities in these domains are $u(t, x)$, $w(t,x)$, and $v(t,x)$, respectively. In the outer layers, the flow is assumed to be slightly shear and can be described using average velocities: $U\approx u$ for $y\in(0,\xi)$, $U\approx w$ for $y\in(Y-\zeta,Y)$. A vortex flow is realized in the mixing layer. Therefore, in addition to the average flow velocity $v$ another variable $q(t,x)$ is introduced. This quantity characterizes the spatial inhomogeneity of the flow (and will be defined below). It is assumed that the fluid motion mainly occurs in the direction of the $Ox$ axis and the ratio of the characteristic transverse scale of the flow $Y$ to the longitudinal scale of $L$ is small $\varepsilon=Y/L \ll 1$. For convenience of further calculations we assume $U\geq 0$.

As a result of averaging of Eqs.~(\ref{eq:SW_2D}) over the channel width with the above-mentioned assumptions, the following system is obtained
\begin{equation} \label{eq:3L}
 \begin{array}{l} \displaystyle 
   \xi_t+(u\xi)_x=-\sigma q, \quad \eta_t+(v\eta)_x=2\sigma q, \quad
   \zeta_t+(w\zeta)_x=-\sigma q, \\[3mm]\displaystyle 
   u_t+(u^2/2+p)_x=-c u^2, \quad w_t+(w^2/2+p)_x=-c w^2, \\[3mm]\displaystyle
   Q_t+\big(u^2\xi+(v^2+kq^2)\eta+w^2\zeta+Yp\big)_x= -c(u^2\xi+(v^2+kq^2)\eta+w^2\zeta), \\[3mm]\displaystyle
   \big(u^2\xi+(v^2+q^2)\eta+w^2\zeta\big)_t+ \big(u^3\xi+(v^2+(1+2k)q^2)v\eta+w^3\zeta+2Qp\big)_x= 
   \\[3mm]\displaystyle \quad\quad\quad 
   =-2c(u^3\xi+(v^2+(1+2k)q^2)v\eta+w^3\zeta)-\kappa\sigma q^3.
 \end{array} 
\end{equation}
Here $Q=u\xi+v\eta+w\zeta$ is the fluid flow rate (referred to the channel thickness $h$); constant $c=c_f/h$; non-negative constants $\sigma$ and $\kappa$ are the empirical parameters responsible for mixing between layers and energy dissipation; the coefficient $k$ takes values 0 and 1, which corresponds to Models I and II. In both cases, Eqs.~(\ref{eq:3L}) form a closed system for determining the width of the layers $\xi$, $\zeta$, and $\eta=Y-\xi-\zeta$, the average fluid velocity in these layers $u$, $w$, and $v$, the pressure $p$, and the variable $q\geq 0$. Let us proceed to the derivation of Eqs.~(\ref{eq:3L}).

\subsection{Derivation of Model II} 

We perform the following scaling in Eqs.~(\ref{eq:SW_2D}) and (\ref{eq:energy-gen})  
\[ t\to\varepsilon^{-1} t, \quad x\to\varepsilon^{-1} x, \quad 
   V\to\varepsilon V, \quad c_f\to\varepsilon c_f\,. \]
Then we neglect all terms of the order of $\varepsilon^2$. As a result, we obtain the equations of the horizontal-shear flow of a thin fluid layer in a closed channel 
\begin{equation} \label{eq:HSF}
 \begin{array}{l} \displaystyle
   U_t+(U^2+p)_x+(UV)_y=-c\,U^2, \quad p_y=0, \quad U_x+V_y=0; \\[3mm]\displaystyle
   \frac{\partial }{\partial t} \bigg(\frac{U^2}{2}\bigg)+ 
   \frac{\partial }{\partial x} \bigg(\bigg(\frac{U^2}{2}+p\bigg)U\bigg)+  
   \frac{\partial }{\partial y} \bigg(\bigg(\frac{U^2}{2}+p\bigg)V\bigg) = -c\,U^3\,,
 \end{array}
\end{equation}
The last equation in (\ref{eq:HSF}) is a consequence of the first three ones. We supplement system~(\ref{eq:HSF}) with boundary conditions~(\ref{eq:BC}).

Let us assume that the velocity profile $U(t,x,y)$ satisfies the condition
\begin{equation} \label{eq:Uy}
  U_y =
   \left\{
   \begin{array}{ll}
     O(\varepsilon^\beta),     & \quad y \in (0, \xi) \cup (Y-\zeta, Y) \\[2mm]
     O(\varepsilon^{\gamma}), & \quad y \in (Y+\xi, Y-\zeta)
   \end{array}
   \right. 
\end{equation}
with the parameters $\beta\in (1,2]$ and $\gamma\in (0,1)$. To describe a three-layer flow, we introduce the averaged velocities in the layers and the root-mean-square deviation (shear velocity~$q$) in the intermediate layer:
\begin{equation} \label{eq:aver-vel} 
  u=\frac{1}{\xi}\int\limits_0^\xi U\,dy, \quad 
  w=\frac{1}{\zeta}\int\limits_{Y-\zeta}^{Y} U\,dy, \quad
  v=\frac{1}{\eta}\int\limits_\xi^{\xi+\eta} U\,dy, \quad 
  q^2=\frac{1}{\eta}\int\limits_\xi^{\xi+\eta} (U-v)^2\,dy\,. 
\end{equation}

The process of fluid involvement in the vortex interlayer is assumed to be symmetric and is taken into account by kinematic conditions at the internal boundaries
\begin{equation} \label{eq:kin-cond} 
 \begin{array}{l} \displaystyle
  \xi_t+U\xi_x-V\big|_{y=\xi}=-\sigma q, \quad (\xi+\eta)_t+U(\xi+\eta)_x-V\big|_{y=\xi+\eta}=\sigma q.
 \end{array}
\end{equation}
This means that the rate of fluid entrainment into the vortex interlayer is proportional to the velocity of ``large eddies'' generated in the vicinity of the interface between the layers due to shear instability \cite{ChL_2019, GLCh16}. According to \cite{LT00}, the proportionality coefficient $\sigma$ is approximately equal to $0.15$. 

In averaging Eqs.~(\ref{eq:HSF}) over the channel width the necessity of approximate calculation of the integrals of the functions $U^n$ ($n=2,3$) with respect to the variable $y$ arises. The following estimates are valid for weakly-shear layers:
\[ \int\limits_0^\xi U^n\,dy=\xi u^n+O(\varepsilon^{n\beta}), \quad 
   \int\limits_{Y-\zeta}^Y U^n\,dy=\zeta w^n+O(\varepsilon^{n\beta}) \quad (1<\beta\leq 2). \]
As terms of the order of $\varepsilon^2$ are ignored in deriving model (\ref{eq:HSF}), the velocity $U$ is replaced in the course of averaging the equations of motion in the weakly shear layers by the corresponding averaged values of $u$ and $w$ determined by formulas (\ref{eq:aver-vel}). Using the obvious identity $U=v+(U-v)$, we calculate the integrals in the intermediate mixing layer:
\begin{equation} \label{eq:int} 
  \int\limits_\xi^{\xi+\eta} U^2\,dy=(v^2+q^2)\eta, \quad 
  \int\limits_\xi^{\xi+\eta} U^3\,dy=(v^2+3q^2)v\eta+C_3, \quad 
  C_3=\int\limits_\xi^{\xi+\eta} (U-v)^3\,dy\,. 
\end{equation}
Condition (\ref{eq:Uy}) implies the estimate $C_3=O(\varepsilon^{3\gamma}) \ll\eta v q^2= O(\varepsilon^{2\gamma})$, $0<\gamma<1$. Therefore, the value of $C_3$ is small compared to other terms and can be omitted~\cite{Tesh_2007}. 

The fulfilment of conditions (\ref{eq:Uy}) imposed on the velocity profile $U(t,x,y)$ is sufficient to require at the initial moment of time $t=0$. Indeed, by virtue of (\ref{eq:HSF}), the variable $U_y$ satisfies the equation
\[ \frac{dU_y}{dt}=-2cUU_y \quad\quad \bigg(\frac{d}{dt}=\frac{\partial}{\partial t}+
   U\frac{\partial}{\partial x} +V\frac{\partial}{\partial y}\bigg)\,, \]
and, consequently, $|U_y|$ decreases with time.

With allowance for the adopted assumptions, as a result of averaging Eqs.~(\ref{eq:HSF}) over the channel width and using boundary conditions~(\ref{eq:BC}) and (\ref{eq:kin-cond}), we obtain system~(\ref{eq:3L}) for $k=1$ (Model II). The first three equations of system~(\ref{eq:3L}) express the mass balance in the layers. Since $\xi+\eta+\zeta=Y={\rm const}$ then the flow rate does not depend on the cross section, i.e. $Q=Q(t)$. The fourth and fifth equations are the local momentum conservation laws in the outer layers, the sixth and seventh equations are the total momentum and energy conservation laws. The last term in the energy equation with the factor $\kappa$ does not directly follow from the averaging over the channel width of the last equation in (\ref{eq:HSF}). It is added to take into account the energy dissipation. According to \cite{LT00}, the numerical value of the empirical parameter $\kappa$ lies in the interval from 2 to 8.

Governing equations (\ref{eq:3L}) imply the fulfilment of the conditions
\[ U\big|_{y=\xi}=u, \quad U\big|_{y=\xi+\eta}=w. \]
They are the consequence of the compatibility of the equations of balance of mass, momentum and energy in weakly-shear layers. A similar approach for determining the velocity at the interface was used in \cite{ChL_2019, GLCh16}. Indeed, averaging Eqs.~(\ref{eq:HSF}) over the width of one of the outer layers yields the system
\begin{equation} \label{eq:mom-en-1}  
 \begin{array}{l} \displaystyle 
   \xi_t+(u\xi)_x=-\sigma q, \quad 
   (u\xi)_t+(u^2 \xi)_x+\xi p_x+c u^2\xi+\sigma qU\big|_{y=\xi}=0,  \\[3mm]\displaystyle    
   (u^2\xi)_t+(u^3\xi+2u\xi p)_x-2p(u\xi)_x+2c u^3\xi+\sigma q U^2\big|_{y=\xi}=0.
  \end{array}  
\end{equation}
Averaging over another weakly-shear layer gives the same equations up to the replacement of $u$, $\xi$ and $U|_{y=\xi}$ with $w$, $\zeta$ and $U|_{y=\xi+\eta}$. It is easy to show that the compatibility of system~(\ref{eq:mom-en-1}) for $\sigma q\neq 0$ is ensured by the condition $(U-u)^2|_{y=\xi}=0$.

\subsection{Derivation of Model I}

At large Reynolds numbers, vortex motions of various scales arise as a result of the development of shear instability. Let a certain scale be chosen and a procedure for averaging the unknown variables be constructed. Then the velocity field and pressure can be represented as
\begin{equation} \label{eq:turb-var} 
  U=\overline{U}+U', \quad V=\overline{V}+V', \quad p=\overline{p}+p',
\end{equation}
where $\overline{U}\geq 0$, $\overline{V}$ and $\overline{p}$ are the average velocity components and pressure; $U'$, $V'$ and $p'$ are the corresponding fluctuations relative to the average variables such that $\overline{U'}= \overline{V'}= \overline{p'}=0$. We substitute representation~(\ref{eq:turb-var}) into system~(\ref{eq:SW_2D}) and energy equation (\ref{eq:energy-gen}). After averaging over the selected scale, we obtain the Reynolds equations:
\begin{equation} \label{eq:2D-turb} 
 \begin{array}{l}\displaystyle 
  \overline{U}_t+ \big(\overline{U}^2+P\big)_x+ 
  \big(\overline{U}\,\overline{V}-\tau\big)_y= 
  -c\overline{U}\sqrt{\overline{U}^2+\overline{V}^2}, \\[3mm]\displaystyle
  \overline{V}_t+ \big(\overline{U}\,\overline{V}-\tau\big)_x+ 
  \big(\overline{V}^2+P\big)_y= -c\overline{V}\sqrt{\overline{U}^2+\overline{V}^2}, \quad
  \overline{U}_x+\overline{V}_y=0,  \\[3mm]\displaystyle
  \frac{\partial }{\partial t}\bigg(\frac{\overline{U}^2+\overline{V}^2+\hat{q}^2}{2}\bigg)+ 
  \frac{\partial }{\partial x}\bigg(\bigg(\frac{\overline{U}^2+\overline{V}^2+\hat{q}^2}{2}+
  P\bigg)\overline{U}-\tau\overline{V}\bigg)+ \\[4mm]\displaystyle
  \quad\quad\quad +\frac{\partial }{\partial y}\bigg(\bigg(\frac{\overline{U}^2+\overline{V}^2+\hat{q}^2}{2}+
  P\bigg)\overline{V}-\tau\overline{U}\bigg)=
  -c\big(\overline{U}^2+\overline{V}^2\big)^{3/2} -\omega \hat{q}^2.  
 \end{array}
\end{equation}   
Here $\tau=-\overline{U'V '} $ is the Reynolds shear stress; $P=\overline {p}+\overline{{U'}^2}$ is the total pressure; $\hat{q}^2=\overline{{U'}^2}+\overline{{V'}^2}$ is the specific kinetic energy of the fluctuating motion. On the right-hand sides of Eqs.~(\ref{eq:SW_2D}), the true velocities are replaced by their average values without fluctuating terms. In deriving Eqs.~(\ref{eq:2D-turb}) an assumption is made about the isotropy of the fluctuating motion, i.e. $\overline{{U'}^2}=\overline{{V'}^2}$. Moreover, the terms $\overline{U'p'}$, $\overline{V'p'}$ and $\overline{{U'}^n{V'}^{3-n}}$ ($n=0,...,3$) having a higher order of smallness compared to $\hat{q}^3$ are neglected. The last term in the fourth equation~(\ref{eq:2D-turb}) is added to take into account the outflow of energy into small-scale motion. The characteristic frequency $\omega$ is responsible for the energy dissipation rate of turbulent motion. In view of condition~(\ref{eq:BC}) on the side walls of the channel $y=0$ and $y=Y$ the velocity component $\overline{V}$ vanishes.

In a developed turbulent flow the following dependence for the Reynolds shear stress was experimentally confirmed in \cite{Townsend_1956}
\begin{equation} \label{eq:tau}  
  \tau=\sigma \hat{q}^2\, {\rm sign}\,\overline{U}_y.
\end{equation} 
Eqs.~(\ref{eq:2D-turb}), (\ref{eq:tau}) are the closed system for determining the functions $\overline{U}$, $\overline{V}$, $P$ and $\hat{q}$. Taking into account that the fluid moves mainly along the $Ox$ axis, we proceed to the approximation of a long channel ($\varepsilon^2\to 0$). We assume here that the fluctuating components $\overline{{U'}^2}$, $\overline{U'V'}$ and the variable $\hat{q}^2$ are of the same order. Then the following scaling 
\begin{equation} \label{eq:scaling} 
 \begin{array}{l}\displaystyle  
  y\to\varepsilon y, \quad t\to\varepsilon^{-1/2}t, \quad \overline{U}\to\varepsilon^{1/2}\overline{U}, \quad \overline{V}\to\varepsilon^{3/2}\overline{v}, \\[2mm]\displaystyle 
  P\to\varepsilon P, \quad \hat{q}\to\varepsilon^{1/2}\hat{q}, \quad \omega\to \varepsilon^{-1/2} \omega
 \end{array}
\end{equation} 
in Eqs.~(\ref{eq:2D-turb}) and neglecting the leading terms in powers of the small parameter $\varepsilon$ yields
\begin{equation} \label{eq:2D-turb-LW} 
 \begin{array}{l}\displaystyle 
  \overline{U}_t+ \big(\overline{U}^2+P\big)_x+ \big(\overline{U}\,\overline{V}- \varepsilon^{-1}\tau\big)_y= 
  -c\overline{U}^2, \quad P_y=0, \quad \overline{U}_x+\overline{V}_y=0, \\[4mm]\displaystyle
  \bigg(\frac{\overline{U}^2+\hat{q}^2}{2}\bigg)_t+
  \bigg(\Big(\frac{\overline{U}^2+\hat{q}^2}{2}+P\Big)\overline{U}\bigg)_x+ 
  \bigg(\Big(\frac{\overline{U}^2+\hat{q}^2}{2}+P\Big)\overline{V}-
  \frac{\tau\overline{U}}{\varepsilon} \bigg)_y= \\[4mm]\displaystyle
  \hspace{8cm} =-c\overline{U}^3- \varepsilon^{-1}\omega \hat{q}^2,
 \end{array}
\end{equation}  
where $\tau$ is defined by formula (\ref{eq:tau}). Following \cite{LT00}, we assume $\omega=\sigma\kappa \hat{q}/\eta$. It should be noted that in scaling (\ref{eq:scaling}) the value $\varepsilon$ has the same order of smallness as the parameter $\sigma$, therefore, the terms $\sigma/\varepsilon$ are kept in Eqs.~(\ref{eq:2D-turb-LW}). Further, in Eqs.~(\ref{eq:2D-turb-LW}) we take $\varepsilon=1$.

Let us average Eqs.~(\ref{eq:2D-turb-LW}) over the channel width taking into account a three-layer representation of the flow (Fig.~\ref{fig:fig_1}). We assume that in the outer weakly-shear layers of width $\xi(t,x)$ and $\zeta(t,x) $ there is no kinetic energy of fluctuating motion ($\hat{q}^2=0$), and the flow velocity $\overline{U}$ is equal to $u(t,x)$ and $w(t,x)$, respectively. In the intermediate mixing layer of width $\eta(t,x)$ we replace velocity $\overline{U}(t,x,y)$ by its average value $v(t,x)$ and denote $q^2(t,x)$ the average value of the function $\hat{q}^2(t,x,y)$. We recall that upon obtaining Eqs.~(\ref{eq:2D-turb}) (and (\ref{eq:2D-turb-LW})), a certain averaging scale is fixed. If, as such a scale, we choose the width of the mixing layer $\eta$, then $\overline{U}=v$. In view of the second equation in (\ref{eq:2D-turb-LW}) the total pressure is $P=p(t,x)$. At the interfaces $y=\xi$ and $y=\xi+\eta$, it is assumed that kinematic conditions~(\ref{eq:kin-cond}) are satisfied (with velocities $\overline{U}$ and $\overline{V} $ instead of $U$ and $V$). As a result of averaging Eqs.~(\ref{eq:2D-turb-LW}) over the channel width, we obtain system~(\ref{eq:3L}) for $k=0$ (Model~I).

\subsection{Transformation of Eqs.~(\ref{eq:3L})} 

For further analysis of the equations of motion and the construction of stationary solutions it makes sense to derive some consequences. By virtue of system~(\ref{eq:3L}) the variables $u-w$, $v$ and $q$ satisfy the equations
\begin{equation} \label{eq:d-cons} 
 \begin{array}{l} \displaystyle  
   (u-w)_t+\Big(\frac{u^2-w^2}{2}\Big)_x=-c(u^2-w^2), \\[3mm]\displaystyle 
   v_t+ vv_x +p_x +2kqq_x+ \frac{kq^2}{\eta}\eta_x= 
   \frac{\sigma q}{\eta}(u-2v+w)-c(v^2+kq^2), \\[3mm]\displaystyle 
   q_t+vq_x+kqv_x=\frac{\sigma}{2\eta}\big((u-v)^2+(w-v)^2-(2+\kappa)q^2\big)-(k+1)cvq=f_q.
 \end{array} 
\end{equation}

Using the sixth equation in system (\ref{eq:3L}) we eliminate $p_x$, replace the energy equation with the differential consequence for the variable $q$, and also express $\eta$ and $v$ in terms of other variables. Then system (\ref{eq:3L}) is transformed to the following equations to determine five unknowns $\xi$, $u$, $\zeta$, $w$ and $q$:
\begin{equation} \label{eq:3L-mod}
 \begin{array}{l} \displaystyle 
   \xi_t+(u\xi)_x=-\sigma q, \quad u_t+\Big(\frac{u^2}{2}-G\Big)_x= c(G-u^2)+ \frac{Q'(t)}{Y},
   \\[3mm]\displaystyle
   \zeta_t+(w\zeta)_x=-\sigma q, \quad w_t+\Big(\frac{w^2}{2}-G\Big)_x= c(G-w^2)+ \frac{Q'(t)}{Y}, \\[3mm]\displaystyle
   q_t+vq_x+kqv_x=f_q.
 \end{array} 
\end{equation}
Here function $f_q$ is defined in the last formula (\ref{eq:d-cons}), the variables $\eta$, $v$ and $G$ are expressed as follows:
\[ \eta=Y-\xi-\zeta, \quad v=\frac{Q(t)-u\xi-w\zeta}{\eta}, \quad 
   G=\frac{u^2\xi+(v^2+kq^2)\eta+w^2\zeta}{Y}\,. \]
The flow rate $Q(t)$ is considered to be given. The pressure $p$ on the rigid-lid is found from the fourth or fifth equation of system (\ref{eq:3L}). As it was mentioned above the coefficient $k=0$ for Model I and $k=1$ in the case of Model II. Eqs.~(\ref{eq:3L-mod}) is used below for an averaged description of the mixing layer. 

\subsection{Characteristics of Eqs.~(\ref{eq:3L-mod})} 

Let us represent Eqs.~(\ref{eq:3L-mod}) in the form
\[ \bU_t+\bA\bU_x=\bF, \]
where $\bU=(\xi, u, \zeta, w, q)^{\rm T}$ is the vector of unknown variables, $\bA$ and $\bF$ is the corresponding matrix of size $5\times 5$ and the right-hand side vector. The eigenvalues $\lambda=\lambda_i$ of matrix $\bA $ determine the propagation velocity of the characteristics. By virtue the last equation in system~(\ref{eq:3L-mod}), Model~I ($k=0$) has the contact characteristic $dx/dt=v$. Model II ($k=1$) has the same characteristic since equation $s_t+vs_x=f_s$ holds on for the variable  $s=q/\eta$ (here the right-hand side term $f_s$ does not contain derivatives of the unknown functions). To determine the others characteristics of system~(\ref{eq:3L-mod}), an algebraic equation of the fourth degree arises
\[ \begin{array}{l} \displaystyle  
    \chi(\lambda)=(u-\lambda)(w-\lambda) \Big((u-\lambda)(w-\lambda)- \frac{2\xi}{Y}(u-v)(w-\lambda)-   
    \frac{2\zeta}{Y}(w-v)(u-\lambda)\Big)- \\[3mm]\displaystyle
    \quad\quad\quad 
    +\frac{\zeta}{Y}\big((w-v)^2-3kq^2\big)(u-\lambda)^2
    +\frac{\xi}{Y}  \big((u-v)^2-3kq^2\big)(w-\lambda)^2 =0.
   \end{array} \]

The case $\xi=\zeta\to Y/2$ corresponds to the initial point of confluence of two flows of the same width. The polynomial $\chi(\lambda)$ is simplified and takes the form
\[ \chi(\lambda)=\big((u-\lambda)^2+(w-\lambda)^2\big)\big((v-\lambda)^2-3kq^2\big)/2. \]
For both models ($k=0$ and $k=1$), equation $\chi(\lambda)=0$ has two complex and two real roots. It is necessary to note that in the framework of Model~I, all real characteristics are contact and propagate with the average velocity $v$ in the intermediate mixing layer. Thus, the system under consideration is not hyperbolic in the vicinity of the confluence of flows. It is associated with the development of the Kelvin--Helmholtz instability at the interface.

\begin{rmk}
As it is shown in \cite{ChL_2019} a similar to model~(\ref{eq:3L}) system of equations describing a mixing layer for horizontal-shear flow with a free surface has, at least, three real characteristics (one contact and two sonic ones). Moreover, under certain conditions this system is hyperbolic, that allows one to use standard numerical methods developed for solving hyperbolic equations. A modification of system (\ref{eq:3L}) for shear flows of a weakly compressible barotropic fluid under a rigid-lid also has real sonic characteristics and consequently can be used to model non-stationary processes in mixing layers.
\end{rmk}

\section{Stationary solutions} 

The steady-state fluid flows in the framework of three-layer model (\ref{eq:3L-mod}) are determined by solving the system of ODE
\[ \begin{array}{l} \displaystyle 
    (u\xi)'=-\sigma q, \quad (w\zeta)'=-\sigma q, \quad uu'-G'= c(G-u^2), \\[3mm]\displaystyle 
    uu'-ww'=-c(u^2-w^2), \quad vq'+kqv'=f_q.
   \end{array} \]
We denote by ``prime'' the derivative with respect to $x$. After some transformations the system is rewritten in the normal form:
\begin{equation} \label{eq:st-3L} 
 \begin{array}{l} \displaystyle  
  w'=\frac{F}{\Delta}, \quad u'=\frac{w}{u}w'-\frac{c}{u}(u^2-w^2), \quad
  \xi'=-\frac{\sigma q+\xi u'}{u}, \\[4mm]\displaystyle 
  \zeta'=-\frac{\sigma q+\zeta w'}{w}, \quad 
  q'=\frac{f_q}{v}-\frac{k}{\eta}\bigg(\frac{2\sigma q^2}{v}+(\xi'+\zeta')v\bigg).
   \end{array} 
\end{equation}
Here
\[ \begin{array}{l} \displaystyle 
    F=c\bigg(\frac{(v^2+kq^2-w^2)\eta}{v^2-3kq^2}+\frac{(u^2-w^2)\xi}{u^2} \bigg)
    -\sigma q\bigg(\frac{u+w-4v}{v^2-3kq^2}+\frac{1}{u}+\frac{1}{w}\bigg)+ \\[4mm]\displaystyle
    \quad\quad\quad +k\frac{2q(\eta f_q-2\sigma q^2)}{v(v^2-3q^2)}, \quad\quad
    \Delta=\bigg(\frac{\xi}{u^2}+\frac{\eta}{v^2-3kq^2}+\frac{\zeta}{w^2}\bigg)w,
   \end{array} \]
the function $f_q$ is the right-hand side in the last equation in (\ref{eq:d-cons}), the coefficient $k$ takes values 0 and 1 (Models I and II, respectively). 

\subsection{Initial section of the mixing layer} 

The equations describing a stationary mixing layer are derived above. To construct the solution, it is necessary to impose the conditions at the point of mixing layer formation, i.e., to find the asymptotic of the stationary solution (\ref{eq:st-3L}) as $\eta\to 0$. Without loss of generality, we assume that $\eta=0$ at $x=0$, and the values of the functions at this point are marked by the zero subscript. Let at $x=0$ the parameters of weakly shear layers $u_0$, $w_0$, $\xi_0$ and $\zeta_0=Y-\xi_0$ be specified (all values are positive). We assume that there are finite limits of the functions $v\to v_0>0$, $q\to q_0>0$ as
$x\to 0$ and find these values.

By virtue of the last two equations in (\ref{eq:d-cons}) and (in the case $k=1$) the second equation in (\ref{eq:3L}) we have the relations
\begin{equation} \label{eq:lim}  
  kq_0^2=(u_0+w_0-2v_0)v_0/2, \quad (u_0-v_0)^2+(w_0-v_0)^2=(2+\kappa)q_0^2. 
\end{equation}
For model I ($k=0$) it follows from the first equation in (\ref{eq:lim}) that $v_0=(u_0+w_0)/2$ and from the second equation in (\ref{eq:lim}) the value of $q_0^2$ is determined. For model II ($k=1$) relations (\ref{eq:lim}) are reduced to the quadratic equation for determining $r=v_0/u_0$
\[ (1-r)^2+(r_0-r)^2=(2+\kappa)(1+r_0-2r)r/2. \] 
Here we denote $r_0=w_0/u_0$. This equation has two real roots $r=r_{1,2}$ for all $r_0>0$ if $\kappa>\kappa_*$, where $\kappa_*\approx 7.657$. From the two roots $r_1$ and $r_2$, we choose the closest to the value of $(1+r_0)/2$. If $r_0$ tends to unity (the velocities of the merging flows $u_0$ and $w_0$ are close), then the region of existence of the quadratic equation solution with respect to the parameter $\kappa$ expands. The indicated restriction on the choice of the dissipation parameter $\kappa$ should be taken into account in calculations according to Model II.

\subsection{Stationary mixing layer} 

We obtain a numerical solution of Eqs.~(\ref{eq:st-3L}) for the parameters of the incoming flow taken from  \cite{Booij_2001} (case A): $u_0=0.118$~m/s, $w_0=0.238$~m/s. The calculations are performed for a straight channel having 16~m long and 3~m wide. Its depth $h$ is equal to 0.052~m. We assume $\xi_0=\zeta_0=1.5$~m,  $c_f=0.003$ and $\kappa=4$. For this and all subsequent tests, we choose $\sigma=0.15$. The solution of the stationary mixing layer problem is not difficult, since it is described by a system of ordinary differential equations. The calculations are carried out in the MATLAB environment using the ode45 function that implements the fourth-order Runge--Kutta method.

\begin{figure}[t]
\begin{center}
\resizebox{1\textwidth}{!}{\includegraphics{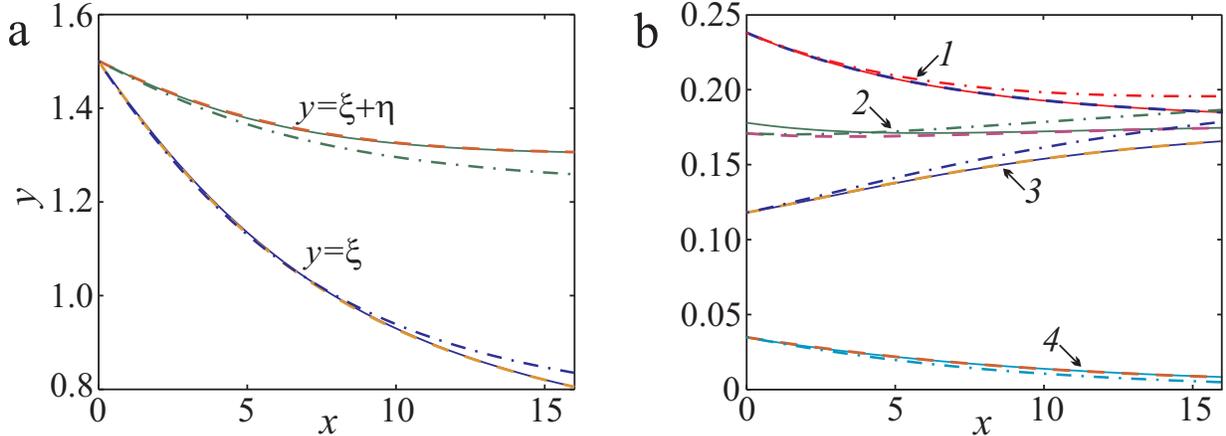}}\\[0pt]
{\caption{Flow parameters in the mixing layer: a --- internal boundaries $y=\xi$ and $y=\xi+\eta$; b --- average fluid velocities in the layers $u$, $v$, $w$ and variable $q$ (curves {\it 1}, {\it 2}, {\it 3} and {\it 4}). Solid and dashed curves are the solution of Eqs.~(\ref{eq:st-3L}) for $k=0$ and $k=1$ (Models I and II), respectively; dash-dotted curves show the solution of equations for the flow with a free surface \cite{ChL_2019}.} \label{fig:fig_2ab}} 
\end{center}
\end{figure}

Let two streams moving with the above-given velocities $u_0$ and $w_0$ merge in the cross section $x=0$, where an intermediate mixing layer is formed. For the chosen flow parameters at $x=0$, formulas~(\ref{eq:lim}) uniquely determine the values $v_0\in (u_0,w_0)$ and $q_0$. For Model~I we have $v_0\approx 0.178$, for Model~II --- $v_0\approx 0.171$; in both cases $q_0\approx 0.035$. The results of calculation using Eqs.~(\ref{eq:st-3L}) are shown in Fig.~\ref{fig:fig_2ab} by solid curves (Model I) and dashed lines (Model II). As can be seen from the graphs, in this example, Models I and II give almost the same result. A slight difference in the calculations for these models is observed only for the velocity $v$ at the initial stage of the mixing layer formation. This is due to the difference in definitions of $v_0$ and $q_0$ by formulas~(\ref{eq:lim}) for $k=0$ and $k=1$. Dashed-dotted lines in Fig.~\ref{fig:fig_2ab} correspond to the similar solution (in the framework of Model II) for the flow in an open channel. The solution is obtained in \cite{ChL_2019} and verified by experimental data~\cite{Booij_2001}. Comparison of the calculating results of the mixing layer evolution for shear flows in open and closed channels is not entirely correct. Nevertheless, at the initial stage of the mixing layer formation, a good agreement between the solutions of the equations describing the flows under a rigid-lid and with a free surface is observed.

\begin{rmk}  
By virtue of Eqs.~(\ref{eq:st-3L}) and (\ref{eq:lim}), the position of the mixing layer boundaries depends on the ratio of the velocities of the merging streams $w_0/u_0$. This means that for any $\alpha>0$ the mixing layer boundaries of the flow with velocities $(u_0,w_0)$ and $(\alpha u_0, \alpha w_0)$ coincide.
\end{rmk}

\begin{rmk} 
Eqs.~(\ref{eq:3L}) (and (\ref{eq:st-3L})) include the empirical parameters $\sigma$ and $\kappa$, which are responsible for the entrainment of fluid from the outer layers into the intermediate mixing layer and energy dissipation. A small variation of these parameters leads to a small change in the solution. The detailed analysis of the influence of the variations $\sigma$ and $\kappa $ for two-layer flows with mixing in open channels is given in \cite{Chesn_Ng_2019}.
\end{rmk}

\subsection{Jet flow in a Hele--Shaw cell} 

The calculation results using Eqs.~(\ref{eq:st-3L}) for Models I ($k=0$) and II ($k=1$) practically coincide if the ratio of the flow velocities in the input section satisfies the inequality $1/4<w_0/u_0<4$. As a rule, for problems of jet flows one of the outer fluid layers is at rest. Therefore, the relation $w_0/u_0 \to 0$ (or $w_0/u_0 \to \infty$). In this case, the determination of $v_0$ and $q_0$ from relations (\ref{eq:lim}) for $k=1$ is possible only for sufficiently large values of the dissipation parameter $\kappa>\kappa_*$. In addition, in the framework of Model II, the value of $v_0$ differs markedly from the half-sum of the outer layers velocities $(u_0+w_0)/2$. For this reason, in such problems to apply Model I is preferable. To avoid the singularity in Eqs.~(\ref{eq:st-3L}), the fluid velocity in the resting layer is assumed to be positive, but close to zero. 

The mixing layer boundaries obtained by the proposed model are in a good agreement with the experimental data and the results of direct numerical simulation performed in \cite{Shestakov_2015, Shestakov_2016} for jet Hele--Shaw flows at the Reynolds numbers of the order of $10^4$. We consider jet flow in a closed channel of the size $L\times Y\times h$ (corresponds to the directions $x\times y\times z$), where  $h=2.4$\,mm, $L=267\,h$ and $Y=200\,h$. The channel is filled with fluid at rest. Through a slot of width $9.6\,h$ located in the centre of the cross-section $x=0$ a fluid with velocity $u_0=3.6$~m/s is supplied~\cite{Shestakov_2016}. Below we use this velocity as a scale for the mixing layer reconstruction. Therefore, in dimensionless variables, the jet velocity is chosen equal to one. To visualize the fluid flow at the boundaries of the inlet section, the colouring fluid is injected using needles. As a result of the development of Kelvin--Helmholtz instability, turbulent mixing layers are formed at the interfaces between the jet and the fluid at rest.

\begin{figure}[t]
\begin{center}
\resizebox{.65\textwidth}{!}{\includegraphics{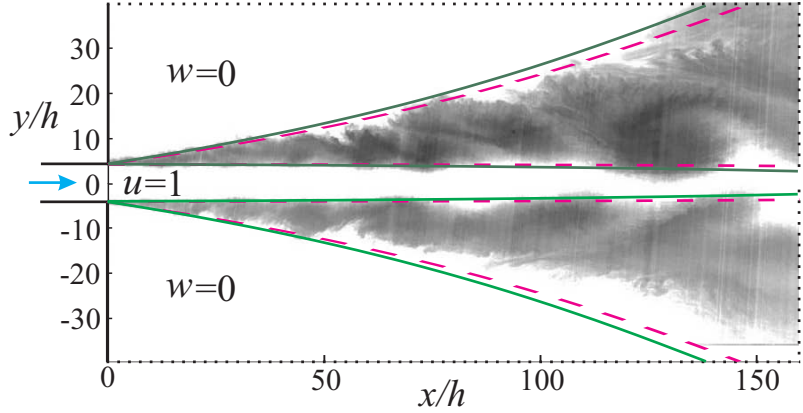}}\\[0pt]
{\caption{Mixing layers arising at the interface between jet flow and fluid at rest in a Hele--Shaw cell. The results of calculating the mixing layer boundaries according to Eqs.~(\ref{eq:st-3L}) (Model I) for $\kappa=6$ (solid curves) and $\kappa=8$ (dashed). Grey-scale picture presents experimental data from \cite{Shestakov_2016} (Fig.~3\,d).}  \label{fig:fig_3}} 
\end{center}
\end{figure}

Let us choose the following parameters $L=200\,h$, $Y=100\,h$, $\xi_0=4.8\,h$, $u_0=1$, $w_0=0.001$ and compute the mixing layer boundaries $y=\xi$ and $y=\xi+\eta$ using Eqs.~(\ref{eq:st-3L}) with $k=0$ (Model I). The channel thickness $h=1$ is chosen as the flow scale. The values of the empirical parameters are $\sigma=0.15$, $\kappa=6$. According to \cite{Dean_1978} the friction coefficient of the two-dimensional rectangular duct flow can be found as 
\begin{equation} \label{eq:c_f}  
  c_f=0.073\,{\rm Re}^{-1/4}\,.
\end{equation}
Therefore, for flows with the Reynolds numbers of the order of $10^4$ considered in \cite{Shestakov_2016}, the friction coefficient is $c_f=0.007$. 

Boundaries of the mixing layer $y=\xi$ and $y=\xi+\eta$ obtained by Eqs.~(\ref{eq:st-3L}), as well as a symmetric about the axis $y=0$ solution corresponding to the conditions $\zeta_0=4.8\,h$, $w_0=1$ and $u_0=0.001$, are shown in Fig.~\ref{fig:fig_3}. Solid curves correspond to the dissipation parameter $\kappa=6$, dashed lines --- $\kappa=8$. The results of these calculations are overlaid on the grey-scaled picture of a turbulent jet presented in \cite{Shestakov_2016} (Fig.~3,\,d). It should be noted that the region of intense mixing mainly propagates into the layer of the fluid at rest and the distance between the internal boundaries of the mixing layers does not decrease until the stability of the jet is lost. As can be seen from Fig.~\ref{fig:fig_3}, Eqs.~(\ref{eq:st-3L}) make it possible to determine the boundaries of intense mixing region in a Hele--Show jet flow without time-consuming calculations on the basis of 2D or 3D equations.

\subsection{Velocity profile and shear stress across the mixing layer} 

Let a solution of Eqs.~(\ref{eq:st-3L}) be constructed in the framework of Model I or II. Below we show that this solution can be applied to determine the velocity profile and Reynolds shear stress across the mixing layer. For this goal, we use stationary equations (\ref{eq:2D-turb-LW}) taking into account the intermittency of the flow (irregularity of the layer boundaries in a turbulent flow), we set (see \cite{Liap_2000} and \cite{LT00}, Ch.~7)
\[ \tau=\sigma q\hat{q}, \quad \omega=\sigma\kappa q/\eta. \]
The functions $q(x)$, $\eta(x)$ and $P=p(x)$ are considered as known ones. They are determined by the solution of stationary Eqs.~(\ref{eq:3L}). Under the above-mentioned assumptions, Eqs.~(\ref{eq:2D-turb-LW}) for steady flows take the form
\begin{equation} \label{eq:2D-st-turb} 
 \begin{array}{l}\displaystyle
  \overline{U}\, \overline{U}_x+\overline{V}\, \overline{U}_y-\sigma q\hat{q}_y+p_x= 
  -c\overline{U}^2, \\[3mm]\displaystyle
  \overline{U}_x+\overline{V}_y=0, \quad 
  \overline{U}\,\hat{q}_x+\overline{V}\,\hat{q}_y-\sigma q\overline{U}_y= -\sigma\kappa\frac{q\hat{q}}{\eta}. 
 \end{array}
\end{equation} 
To analyse this system, it is convenient to pass to the Mises variables $(x,\psi)$, where $\psi$ is the stream function associated with the velocity field by relations $\overline{U}=\psi_y$, $\overline{V}=-\psi_x$.

We denote
\[ \tilde{u}(x,\psi)=\overline{U}(x,y), \quad \tilde{v}(x,\psi)=\overline{V}(x,y), \quad 
   \tilde{q}(x,\psi)=\hat{q}(x,y) \] 
and take into account that 
\[ \overline{U}_x=\tilde{u}_x-\tilde{v}\tilde{u}_{\psi}, \quad 
   \overline{U}_y=\tilde{u}\tilde{u}_{\psi} \]
(derivatives of the functions $\overline{V}$ and $\hat{q}$ are calculated in the similar way). Then Eqs.~(\ref{eq:2D-st-turb}) in the variables $(x,\psi)$ are transformed to the semi-linear hyperbolic system with respect to the variable $x$:
\begin{equation} \label{eq:2D-st-turb-mod}  
 \tilde{u}_x-\sigma q \tilde{q}_\psi=-\frac{p_x}{\tilde{u}}-c\tilde{u}, \quad 
 \tilde{q}_x-\sigma q \tilde{u}_\psi=-\sigma\kappa \frac{q\tilde{q}}{\eta\tilde{u}}\,. 
\end{equation}
The characteristics of system (\ref{eq:2D-st-turb-mod}) are given by the equations $d\psi/dx=\pm\sigma q$. The problem is similar to that considered in \cite{LT00, ChL_2011} for a plane-parallel shear flow in a channel of finite thickness (flow with a pressure gradient). This is formulated for Eqs.~(\ref{eq:2D-st-turb-mod}) in the form of the Goursat problem with data on the characteristics. The difference consists only in the modification of the right-hand side (\ref{eq:2D-st-turb-mod}) related to the consideration of friction. Solution of Eqs.~(\ref{eq:2D-st-turb-mod}) is constructed numerically using a Godunov-type scheme. Then, the velocity profile $\overline{U}(x,y)$ is obtained as a result of integration of the equation $y_\psi=1/\tilde{u}(x,\psi)$ and the transition to the original variables $(x,y)$. 

\begin{figure}[t]
\begin{center}
\resizebox{1\textwidth}{!}{\includegraphics{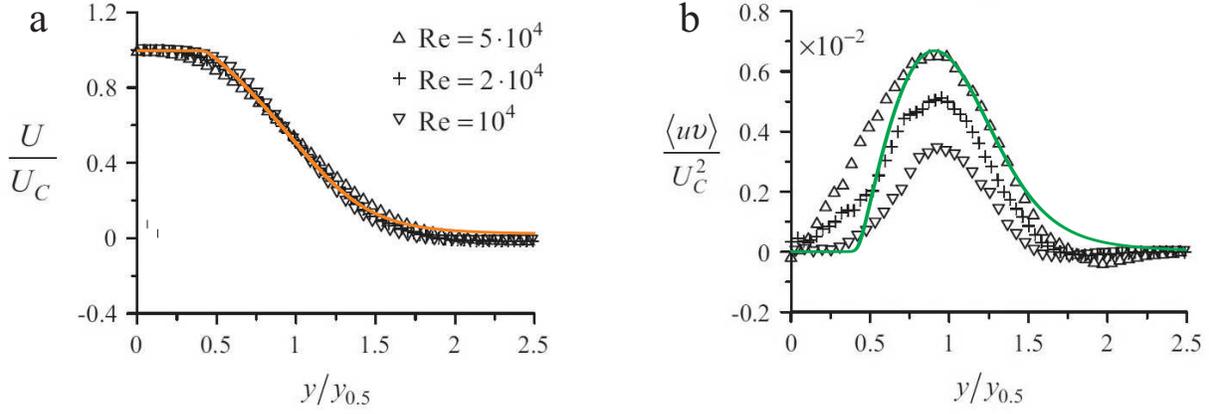}}\\[0pt]
{\caption{The normalized velocity profile $\overline{U}/u$ (a) and Reynolds shear stress $-\sigma q\hat{q}/u^2$ (b) obtained by Eqs.~(\ref{eq:st-3L}) and (\ref{eq:2D-st-turb-mod}) for $x/h=75$ and $d/h=8.75$ are shown by solid curves. These curves are overlaid on Fig.~3 (a) and (b) from \cite{Shestakov_2015}. } \label{fig:fig_4ab}} 
\end{center}
\end{figure}

The results of experimental measurements of the velocity profile and Reynolds shear stress are given in \cite{Shestakov_2015}. This work considers a turbulent jet flow between two parallel plates (separated by a distance $h$) issuing from a rectangular nozzle (with a span-wise size $d$) in the stream-wise direction $x$. Fig.~3 in \cite{Shestakov_2015} presents the measurement results for jet flows with ${\rm Re}=10^4$, $2\cdot 10^4$ and $5\cdot 10^4$ obtained for $d/h=8.75$ at the cross-section $(x-x_0)/h=75$. Here $x_0$ is the virtual origin of the jet ($x_0=0$ for ${\rm Re}=5\cdot 10^4$). The velocity profiles and Reynolds stresses presented in \cite{Shestakov_2015} are normalized to $U_C$ and $U_C^2$, respectively. Here $U_C$ is the centreline velocity in the far jet field taken in the form \cite{Gorin_1998}
\begin{equation} \label{eq:U_C} 
  U_C(x)=\frac{U_{C_0}}{4}\Bigg(\frac{3c_f d}{c_t h}\Bigg)^{1/2}
  \frac{\exp(-c_f(x-x_0)/h)}{(M+1-\exp(-c_f(x-x_0)/h))^{1/2}} 
\end{equation}
with $c_f=0.0075$ and $c_t=0.007$. 

To carry out the calculations based on Eqs.~(\ref{eq:st-3L}) (Model I) and then using Eqs.~(\ref{eq:2D-st-turb-mod}), we choose $h=1$, $Y=100\,h$, $\xi_0=d/2=4.375$, $u_0=1$ and $w_0=0.025$. The empirical parameters are as follows: $\sigma=0.15$, $\kappa=8$ and $c_f=0.0075$. We also assume $x_0=0$. The calculation results (functions $\overline{U}(x,y)$ and $-\sigma q(x)\hat{q}(x,y)$ normalized to the jet velocity $u(x)$ and $u^2(x)$) are shown in Fig.~\ref{fig:fig_4ab} by solid curves at the cross-section $x/h=75$. The value $y_{0.5}$ is chosen so that $\overline{U}=0.5\,u$ for $y=y_{0.5}$. As can be seen from Fig.~\ref{fig:fig_4ab}, the calculation results for the proposed models are in a good agreement with experimental data~\cite{Shestakov_2015}. We note that outside the mixing layer, $\hat{q}=0$ and $\overline{U}$ coincides with the external velocity $u$ (for $y<\xi$) or $w$ (for $y>\xi+\eta$). Since we take $w_0=0.025>0$ to avoid singularity in the governing equations, on the right border of Fig.~\ref{fig:fig_4ab} (a) the calculated velocity profile is slightly higher than the experimental values. The condition $w>0$ is fulfilled everywhere in the flow. The Reynolds shear stress practically coincides with the data from~\cite{Shestakov_2015} for ${\rm Re}=2\cdot 10^4$ in the region $y/y_{0.5}>0.75$. The differences for $y/y_{0.5}<0.5$ can be associated with meandering and penetration of vortices into the central part of the jet. 

\begin{figure}[t]
\begin{center}
\resizebox{0.48\textwidth}{!}{\includegraphics{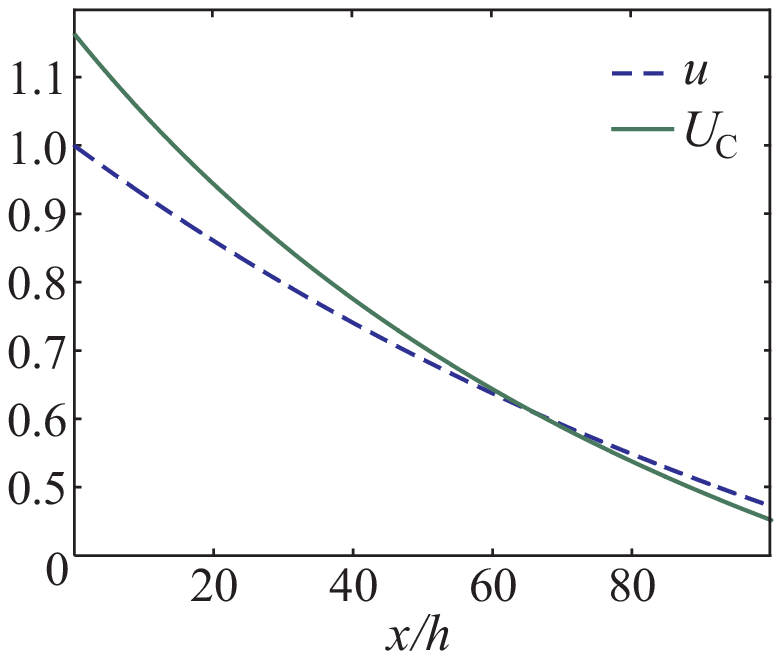}}\hfill
\resizebox{0.48\textwidth}{!}{\includegraphics{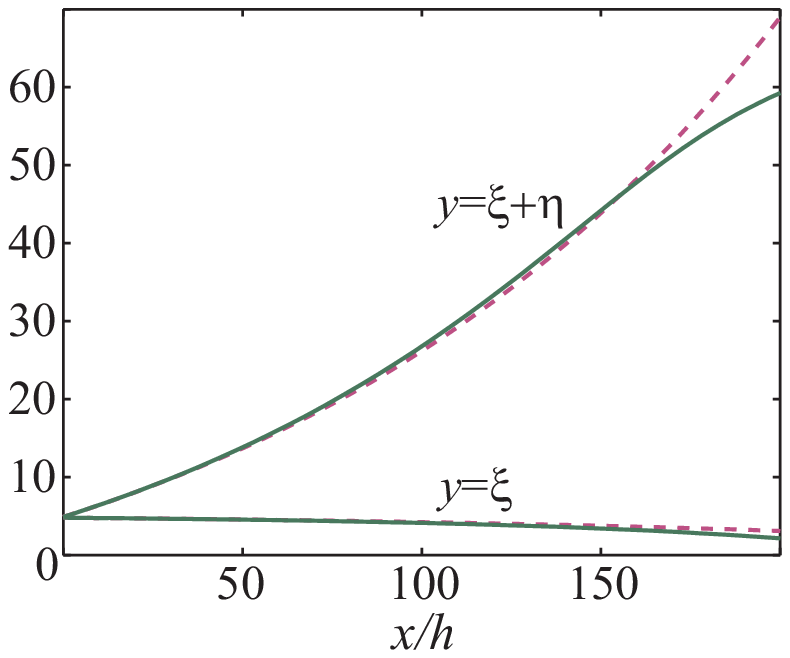}}\\[0pt]
\parbox{0.48\textwidth}{\caption{Stream-wise velocity distribution in the jet $U_C(x)$ given by formula~(\ref{eq:U_C}) (solid curve) and obtained by Eqs.~(\ref{eq:st-3L}) (dashed curve).} \label{fig:fig_5}} \hfill
\parbox{0.48\textwidth}{\caption{The mixing layer boundaries obtained by Eqs.~(\ref{eq:st-3L}) (solid curves) and Eqs.~(\ref{eq:st-3L-p-fr}) (dashed) for $\kappa=6$ and $c_f=0.007$.} \label{fig:fig_6}}
\end{center}
\end{figure}

Fig.~\ref{fig:fig_5} presents the stream-wise velocity in the centre of the jet. It follows from Fig.~\ref{fig:fig_5} that the centreline velocity in the far-field $U_C(x)$ given by the empirical formula~(\ref{eq:U_C}) has the similar behaviour as the velocity $u(x)$ calculated by Eqs.~(\ref{eq:st-3L}) for $x/h>50$. At the considered cross-section $x/h=75$ these values almost coincide. Thus, using the solution of Eqs.~(\ref{eq:st-3L}), we can obtain the transverse distribution of the velocity profile and Reynolds shear stress in the mixing layer on the basis of hyperbolic Eqs.~(\ref{eq:2D-st-turb-mod}).

\section{Hele--Shaw flow without pressure gradient} 

\begin{figure}[t]
\begin{center}
\resizebox{1\textwidth}{!}{\includegraphics{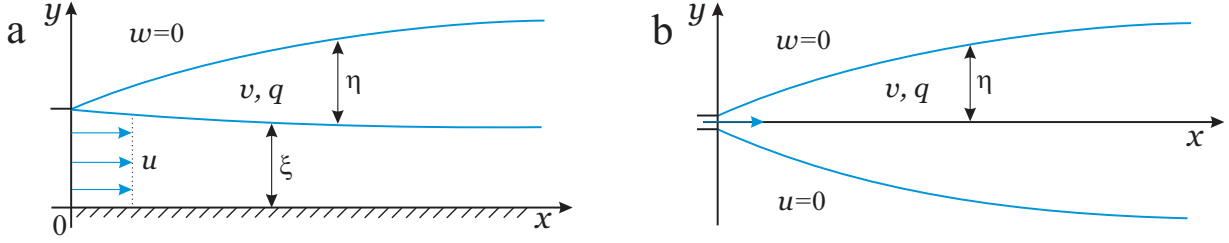}}\\[0pt]
{\caption{Scheme of jet flow in a Hele--Shaw cell without a pressure gradient: a --- mixing layer in a semi-bounded channel; b --- submerged turbulent jet between two parallel planes.} \label{fig:fig_7}} 
\end{center}
\end{figure}

Let us consider the fluid motion in a closed channel of thickness $h$ and unlimited width (the side wall $y=Y$ is considered to be infinitely distant). It corresponds to a flow without a pressure gradient. Further we show that such reduced models can be used to describe at least the initial stage of the stationary mixing layer (Fig.~\ref{fig:fig_6}). As before, we assume that the channel is occupied by the fluid at rest. Through a slot of width $\xi_0$ at $x=0$ a fluid with constant velocity $u_0>0 $ is injected into the channel. The flow scheme is shown in Fig.~\ref{fig:fig_7}\,(a). Taking $w=0$ and $p=0$ in Eqs.~(\ref{eq:3L}), we obtain the following system for determining unknown functions $\xi$, $u$, $\eta$, $v$ and $q$:
\begin{equation} \label{eq:3L-p-fr}
 \begin{array}{l} \displaystyle 
   \xi_t+(u\xi)_x=-\sigma q, \quad \eta_t+(v\eta)_x=2\sigma q, \quad u_t+uu_x=-cu^2, \\[3mm]\displaystyle
   (u\xi+v\eta)_t+\big(u^2\xi+v^2\eta\big)_x= -c(u^2\xi+v^2\eta), \\[3mm]\displaystyle
   \big(u^2\xi+(v^2+q^2)\eta\big)_t+ \big(u^3\xi+(v^2+q^2)v\eta\big)_x= 
   -2c(u^3\xi+(v^2+q^2)v\eta)-\kappa\sigma q^3.
 \end{array} 
\end{equation}
Here we choose $k=0$ (i.e., Model I) to describe Hele--Shaw jet flows. The consequence of system (\ref{eq:3L-p-fr}) are Eqs.~(\ref{eq:d-cons}). In these equations we should put $w=0$ and $p=0$. According to the above-mentioned consequences, stationary equations~(\ref{eq:3L-p-fr}) take the form
\begin{equation} \label{eq:st-3L-p-fr}
 \begin{array}{l} \displaystyle 
   \xi'=c\xi-\frac{\sigma q}{u}, \quad \eta'=c\eta+\frac{\sigma q}{v}\Big(4-\frac{u}{v}\Big), \quad   v'=-cv+\frac{\sigma q}{v}(u-2v), \\[3mm]\displaystyle  q'=-cq+\frac{\sigma}{2v\eta}\Big((u-v)^2+v^2-(2+\kappa)q^2\Big)\,,
 \end{array} 
\end{equation}
where $u=u_0\exp(-2cx)$. At the starting point of the mixing layer formation $x=0 $ due to conditions (\ref{eq:lim}), we have $v_0=u_0/2$,  $q_0=u_0/\sqrt{2(2+\kappa)}$. It should be pointed out that the presence of two integrals
\[ u^2\xi+v^2\eta=(u_0^2\xi_0)\exp(-cx), \quad u\xi+v\eta/2=u_0\xi_0 \] 
allows one to reduce Eqs.~(\ref{eq:st-3L-p-fr}) to a system of two ODEs. 

Boundaries of the mixing layer $y=\xi$ and $y=\xi+\eta$) obtained by Eqs.~(\ref{eq:st-3L-p-fr}) for $u_0=1$, $\xi_0=4.8$, $h=1$, $c_f=0.007$, $\sigma=0.15$ and $\kappa=6$ are shown in Fig.~\ref{fig:fig_6} (dashed curves). For the calculation according to Eqs.~(\ref{eq:st-3L}) (Model I) we additionally put $Y=100$ and $w_0=0.001$. The result of the calculations is shown in the same figure (solid curves). As we can see, at a distance of the order of 160\,$h$ from the input section of the channel, the mixing layer  boundaries practically coincide for the considered models. Therefore, Eqs.~(\ref{eq:st-3L-p-fr}) without a pressure gradient can be used to describe (at least) the initial stage of the mixing layer development for jet flows in a Hele--Shaw cell. 

\subsection{Turbulent jet in a Hele--Shaw cell} 

We consider a quasi-two-dimensional plane turbulent jet discharged from a slot of width $d$ into a fluid confined between two relatively close plane parallel walls with gap $O(d)$. Let us try to describe the averaged boundaries of the jet in the framework of stationary solution of Eqs.~(\ref{eq:3L-p-fr}) without taking into account the pressure gradient. The fluid velocity in both outer layers is equal to zero. The sketch of the flow
 is shown in Fig.~\ref{fig:fig_7}\,(b). Assuming symmetry with respect to the centre line of the channel $y=0$, stationary equations~(\ref{eq:3L-p-fr}) can be written as
\[ (v\eta)'=\sigma q, \quad (v^2\eta)'=-cv^2\eta, \quad ((v^2+q^2)v\eta)'=-2c(v^2+q^2)v\eta-\sigma\kappa q^3. \]
Let us transform these equations to the normal form
\begin{equation} \label{eq:st-jet} 
  \eta'=c\eta+\frac{2\sigma q}{v}, \quad v'=-cv-\frac{\sigma q}{\eta}, \quad 
  q'=-cq+\frac{\sigma}{2v\eta}\big(v^2-(1+\kappa)q^2\big). 
\end{equation}
If we neglect the friction ($c=c_f/h=0$), then Eqs.~(\ref{eq:st-jet}) have the following exact solution
\begin{equation} \label{eq:st-as} 
  \eta=B(x-x_0), \quad v=A(x-x_0)^{-1/2}, \quad q=C(x-x_0)^{-1/2}, 
\end{equation}
where $A$ is the positive constant, $B=2\sigma/\sqrt{\kappa-1}$ and $C=A/\sqrt{\kappa-1}$. Without loss of generality, we take $A=1$ and, as before, we choose $\sigma=0.15$.

\begin{figure}[t]
\begin{center}
\resizebox{.75\textwidth}{!}{\includegraphics{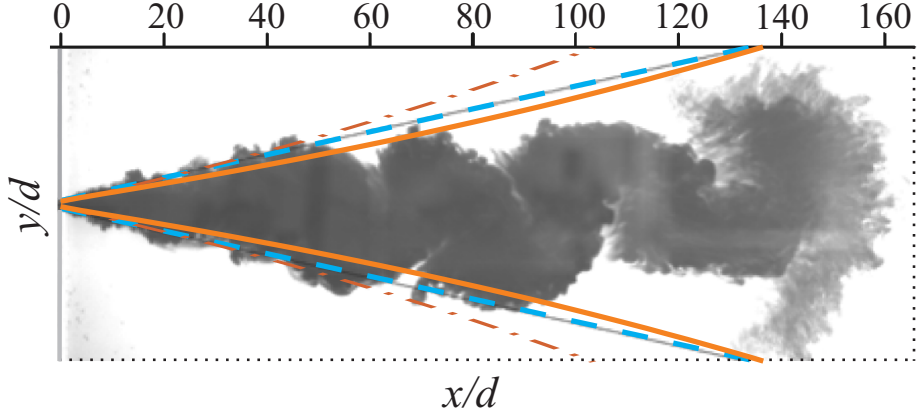}}\\[0pt]
{\caption{Solutions of Eqs.~(\ref{eq:st-jet}) obtained for $\kappa=5$, $c_f=0.0093$ (solid curves),  $\kappa=2.8618$, $c_f=0$ (dashed) and $\kappa=2.8618$, $c_f=0.0093$ (dash-doted curves). Grey-scale picture presents a dyed jet rising in the tank (Fig. 9\,(a) in \cite{Landel_2012b}).} \label{fig:fig_8}} 
\end{center}
\end{figure}

Theoretical and experimental study of steady quasi-two-dimensional plane turbulent jet flows was performed in \cite{Landel_2012a, Landel_2012b}. The self-similar character of the jet spreading at the initial stage is reported in these works. Let us compare the results of calculations of the averaged jet width according to Eqs.~(\ref{eq:st-jet}) with the experimental data \cite{Landel_2012b} obtained in a Hele--Shaw cell of thickness $h=2\,d$, length and width $200\,d$. For the jet flow considered in \cite{Landel_2012b} (see Fig.~9\,(a) in the cited work), the half-spreading angle of the jet is as follows $\theta=12.4^\circ$. This means that in formula~(\ref{eq:st-as}) constant $B$ is equal to $\tan\,\theta\approx 0.2199$. Because the width of the input section is $d$, so that $\eta(0)=d/2$. Hence, we find the virtual origin of the self-similar jet $x_0=-d/(2B)$, as well as the values $v(0)=\sqrt{2B/d}$ and $q(0)=Bv(0)/(2\sigma)$. Due to the relationship between $B$, $\sigma$ and $\kappa$ we define the value of $\kappa=1+4\sigma^2/B^2\approx 2.8618$. The straight line $\eta=B(x-x_0)$ and symmetrical to it relative to the centre line of the channel $y=0$ are shown in Fig.~\ref{fig:fig_8} by dashed curves. These lines coincide with the average dye edges plotted in \cite{Landel_2012b} (Fig. 9\,(a)). 

Presented in \cite{Landel_2012b} grey-scale picture of a dyed jet rising in the tank was obtained for ${\rm Re}= 3850$. Therefore, according to empirical formula~(\ref{eq:c_f}) proposed in \cite{Dean_1978}, we take the friction coefficient $c_f=0.0093$ to calculate the average width of the turbulent flow using Eqs.~(\ref{eq:st-jet}). Solid curves in Fig.~\ref{fig:fig_8} are the solution of Eqs.~(\ref{eq:st-jet}) with the above pointed data at $x=0$ and the dissipation parameter $\kappa=5$. This solution gives a slightly smaller width of the turbulent flow region compared to the self-similar solution~(\ref{eq:st-as}). Dash-dotted curves in the figure represent the solution of Eqs.~(\ref{eq:st-jet}) obtained for $\kappa=2.8618$ and $c_f=0.0093$. This solution has the correct asymptotic behaviour near the inlet cross-section, but in the far field gives an overestimated value of the average jet width. Thus, comparison with Fig.~9\,(a) in \cite{Landel_2012b} shows that Eqs.~(\ref{eq:st-jet}) allow one to determine the average boundary of a turbulent jet flow in a Hele--Shaw cell.

\section{Conclusion} 

One-dimensional models of the formation and evolution of a mixing layer in shear shallow flows of a homogeneous fluid under a rigid-lid (in a Hele--Shaw cell) are proposed. The construction of these models is based on averaging over the channel width of the two-dimensional Reynolds equations with an additional closure condition (Model I) or a quasi-two-dimensional system of equations of shear fluid flow with internal boundaries (Model II). In both cases, the obtained equations of motion are based on a three-layer representation of the flow, taking into account the fluid entrainment process from the outer regions of weakly-shear flow into the vortex interlayer (Fig.~\ref{fig:fig_1}). The rate of fluid entrainment in the mixing layer is proportional to the ``large billows'' velocity generated in the vicinity of the layer interface. The obtained models are written in a unified way and presented in the form of system of five evolution equations~(\ref{eq:3L-mod}) for determining the layers width and average velocities in them, as well as an additional variable responsible for the mixing process. Both models include two empirical constants $\sigma$ and $\kappa$. The first one is responsible for the mixing intensity and the second one for energy dissipation. For all tests we choose $\sigma=0.15$, while $\kappa$ is varied from 5 to 8. It should be note that for shear flows in open channels~\cite{GLCh16, LCh_2014, ChL_2019} more precise coincidence with experimental data is observed for $\kappa \in [3,4] $ and the same parameter $\sigma$. 

The main attention in this paper is paid to the study of stationary solutions. The governing equations are transformed to normal form~(\ref{eq:st-3L}). Stationary solution to the problem of a mixing layer in shallow water under a rigid-lid is obtained and the comparison is made with a similar problem for flows in an open channel (Fig.~\ref{fig:fig_2ab}). If the ratio of velocities in the outer layers differs from unity by less than four times, then Models I and II give almost the same result. Moreover, the initial stage of mixing for flows under a rigid-lid and with a free surface practically coincides. Model I (Eqs.~(\ref{eq:st-3L}) with $k=0$) is more suitable for describing flows in which the velocity of one of the outer layers is close to zero. In this case, according to (\ref{eq:lim}) there is no difficulty in determining the flow parameters at the initial point of the mixing layer formation. This model is verified by comparison with experimental data obtained in \cite{Shestakov_2016} for turbulent Hele--Shaw jet flows. Fig.~\ref{fig:fig_3} shows the results of visualization of the mixing layers when the jet interacts with the fluid at rest as well as the boundaries of the regions of intense mixing calculated using Model I. Further, it is shown that the constructed solution of the averaged equations within a three-layer representation of the flow can be used to obtain the velocity profile and turbulent shear stress across the mixing layer. For this goal, the Reynolds equations in the form~(\ref{eq:2D-st-turb}) are used. In the Mises variables, these equations reduce to a semi-linear hyperbolic system~(\ref{eq:2D-st-turb-mod}). Numerical solution of these equations allow one to obtain the velocity profile and Reynolds shear stress across the mixing layer. The calculation results are verified by the experimental data~\cite{Shestakov_2015} (see Fig.~\ref{fig:fig_4ab} and \ref{fig:fig_5}).

Modification of the derived averaged equations for shear flows without a pressure gradient~(\ref{eq:3L-p-fr}) is considered. This corresponds to quasi-two-dimensional flows with internal boundaries in a channel of unlimited width (see Fig.~\ref{fig:fig_7}). A comparison of the mixing layer boundaries obtained by Eqs.~(\ref{eq:st-3L}) and (\ref{eq:3L-p-fr}) shows, the model without taking into account the pressure gradient is applicable, at least, to describe the initial stage of the mixing layer (Fig.~\ref{fig:fig_6}). In the framework of two-layer equations~(\ref{eq:st-jet}), the boundaries of the turbulent submerged jet in a Hele--Shaw cell are determined. The calculation results are in a good agreement with experimental data~\cite{Landel_2012b} presented in Fig.~\ref{fig:fig_8}. Verification of the models indicates the possibility of their application for an averaged description of jet flows and determination of the boundaries of the large vortex structures formation.

\section*{Acknowledgements}
This work was partially supported by the Russian Foundation for Basic Research (project 19-01-00498). 

\renewcommand\baselinestretch{1}

\end{document}